\documentclass[
prd
,showpacs ,amssymb,nobibnotes,aps,eqsecnum]{revtex4}
\usepackage{graphicx}
\input{epsf}

\usepackage{amsmath,amssymb}
\usepackage{bm}

\newcommand{\dalm}{\kern1pt\vbox{\hrule height 0.9pt\hbox{\vrule width 0.9pt
\hskip 2.5pt\vbox{\vskip 5.5pt}\hskip 3pt\vrule width 0.3pt}\hrule height 0.3pt}
\kern1pt}

\newcommand{\gsim}{\, \raisebox{-0.8ex}{$\stackrel{\textstyle >}{\sim}$ }}
\newcommand{\lsim}{\, \, \raisebox{-0.8ex}{$\stackrel{\textstyle <}{\sim}$ }}

\begin{document}



\title{Slowly Rotating Relativistic Stars in Scalar-Tensor Gravity}

\author{Hajime Sotani} \email{hajime.sotani@nao.ac.jp}
\affiliation{
%
%
Yukawa Institute for Theoretical Physics, Kyoto University, Kyoto 606-8502, Japan
}

\date{\today}

\begin{abstract}
We consider the slowly rotating relativistic stars with a uniform angular velocity in the scalar-tensor gravity, and examine the rotational effect around such compact objects. For this purpose, we derive a 2nd order differential equation describing the frame dragging in the scalar-tensor gravity and solve it numerically. As a result, we find that the total angular momentum is proportional to the angular velocity even in the scalar-tensor gravity. We also show that one can observe the spontaneous scalarization in rotational effects as well as the other stellar properties, if the cosmological value of scalar field is zero. On the other hand, if the cosmological value of scalar field is nonzero, the deviation from the general relativity can be seen in a wide range of the coupling constant. Additionally, we find that the deviation from the general relativity becomes larger with more massive stellar models, which is independent of the cosmological value of scalar field. Thus, via precise observations of astronomical phenomena associated with rotating relativistic stars, one may be possible to probe not only the gravitational theory in the strong-field regime, but also the existence of scalar field.
\end{abstract}

\pacs{04.40.Dg, 04.50.Kd, 04.80.Cc}
%
%
\maketitle
\section{Introduction}
\label{sec:I}

The general relativity proposed by Einstein is the theory describing the gravity. In order to verify the general relativity, many experiments have been done up to now and there is nothing indicating the failure of this theory. However, the most of these experiments are performed in the weak gravitational field, such as the solar system. This means that the gravitational theory in the strong gravitational field is still unconstrained by the astronomical observations. Namely, the gravitational theory might be different from the general relativity in the strong gravitational field. If so, one can probe the gravitational theory by observing the deviation from predictions in the general relativity. Actually, the development of technology will enable us to observe the phenomena related to compact objects with high accuracy. Such observations may be used as a test of the gravitational theory in the strong gravitational field. So far, many attempts and possibilities to test the gravitational theory in the strong gravitational field have been suggested \cite{Will1993,Will2001,Psaltis2008,Paul2008,Sotani2009a,Sotani2010}.
Hopefully the reliable gravitational theory even in the strong gravitational field will be observationally revealed in the near future.

The scalar-tensor gravity is a natural extension of the general relativity, and one of the simplest alternative gravitational theory, where the gravity is mediated by long-range scalar fields in addition to the usual tensor field in the general relativity \cite{Will1993,Damour1992}. The scalar field plays an essential role in explaining the accelerated expansion phases of the universe, such as inflation scenario. Additionally, the scalar-tensor theory in gravity can be obtained from the low energy limit of string and/or other gauge theories. In the scalar-tensor gravity, one adopts the basic assumption that the scalar and gravitational fields, $\varphi$ and $g_{*\mu\nu}$, are coupled to matter via an effective metric defined as $\tilde{g}_{\mu\nu}=A^2(\varphi)g_{*\mu\nu}$, where the function form of the conformal factor is still unclear. The Brans-Dicke theory is the simplest version of the scalar-tensor gravity, which assumes that $A(\varphi)=\exp(\alpha\varphi)$ \cite{BD}. The parameter in the conformal factor, $\alpha$, can be often connected to the Brans-Dicke parameter, $\omega_{\rm BD}$, as $\alpha^2=(2\omega_{\rm BD}+3)^{-1}$, and the expected deviation from the general relativity becomes of the order $\alpha^2$ \cite{Damour1993}. The experiments in the solar system set severe constraints in the parameter, i.e., $\omega_{\rm BD}\gsim 40000$ or $\alpha < 10^{-5}$ \cite{EF2005}. On the other hand, Damour and Esposito-Farese adopted the different function form of the conformal factor, i.e.,  $A(\varphi)=\exp(\alpha\varphi+\beta\varphi^2/2)$. Then, they showed that the massive neutron stars in the scalar-tensor gravity can significantly deviate from the expectations in the general relativity \cite{Damour1993,Damour1996}. Especially, they pointed out that the sudden deviation in the neutron star properties from the predictions in the general relativity is possible for the specific values of the coupling constants, even if the value of $\alpha$ is considerably small. This phenomenon is referred to as ``spontaneous scalarization." Subsequently, Harada studied more details with the technique of catastrophe theory, and found that the spontaneous scalarization is possible for $\beta\lsim -4.35$ \cite{Harada1998}. Recently, Freire {\it et al.} have ruled out the parameter range of $\beta < -5$ with the observations of the pulsar white dwarf binary \cite{Freire2012}. This constraint is quite strict, but we still have a little chance to see the spontaneous scalarization. So, we focus on the range of $\beta \ge-5$ in this article.

Until now, several possibilities to distinguish the scalar-tensor gravity from the general relativity by using astronomical observations are suggested, e.g.,  with the redshift lines of the X- and $\gamma$-rays radiated from the surface of neutron stars \cite{DeDeo2003} and with the spectrum of gravitational waves emitted from neutron stars \cite{Sotani2004,Sotani2005}. In this article, we examine the different approach to probe the gravitational theory. That is, we focus on the rotational effect around neutron stars, where we especially consider the slowly rotating neutron stars with a uniform angular velocity \cite{rotationinBD}. The similar analysis in the general relativity has been originally done by Hartle \cite{Hartle1967}, and subsequently a lot of studies associated with the stellar rotation have been examined. For instance, taking into account the rotational effect, the additional oscillation family referred as $r$-modes can be excited, and one might be possible to obtain the stellar information from their spectrum \cite{Erich2009,Erich2011}. In this sense, this article could become a first step to consider the rotational effect around relativistic stars in more complicated system in the scalar-tensor gravity. We should remark that the similar analysis in the tensor-vector-scalar (TeVeS) theory has been done \cite{Sotani2010}. However, in such an analysis, they did not consider the dependence on the scalar coupling, because in TeVeS the dependence on the vector coupling is stronger than that on the scalar coupling. In addition, for simplicity, they did not consider the dependence on the cosmological value of scalar field. So, in this article, we will see the dependence on the scalar coupling as well as the dependence on the cosmological value of scalar field, although the scalar-tensor gravity is one of the specific case in the TeVeS.

This article is organized as follows. In the next section, we mention the fundamental parts of the scalar-tensor gravity and the equilibrium of non-rotating relativistic stars in the scalar-tensor gravity. In Sec. \ref{sec:III}, we derive the differential equation describing the frame dragging in the scalar-tensor gravity. Additionally, we numerically show the rotational effect in this section. Then, we make a conclusion in Sec. \ref{sec:IV}. In this article, we adopt the unit of $c=G=1$, where $c$ and $G$ denote the speed of light and the gravitational constant, respectively, and the metric signature is $(-,+,+,+)$.

\section{Stellar Models in Scalar-Tensor Gravity}
\label{sec:II}
\subsection{Scalar-Tensor Gravity}
\label{sec:II-1}

The scalar-tensor gravity is a natural extension of the general relativity, where the gravity is mediated not only by metric tensor but also by a massless scalar field $\varphi$. The total action of such a gravitational theory is given by \cite{Damour1992}
\begin{equation}
 S = \frac{1}{16\pi G_*}\int\sqrt{-g_*}\left(R_*-2g_*^{\mu\nu}\varphi_{,\mu}\varphi_{,\nu}\right)d^4x 
     + S_m\left[\Psi_m,\, A^2(\varphi)g_{*\mu\nu}\right],
\end{equation}
where $G_*$ is the bare gravitational coupling constant, $g_{*\mu\nu}$ is the Einstein metric, $R_*$ is the Ricci scalar constructed with $g_{*\mu\nu}$, $\Psi_m$ represents matter fields collectively, and $S_m$ denotes the matter action, respectively. It is known that the formulation of field equations is better in ``Einstein frame" described by the Einstein metric. But a test particle moves on the geodesic in ``Jordan frame" described by the metric $\tilde{g}_{\mu\nu}$, which is defined as $\tilde{g}_{\mu\nu} = A^2(\varphi)\,g_{*\mu\nu}$ \cite{Damour1992}. So, the Jordan frame is often referred to as physical frame. Hereafter, the quantities with asterisk are related to the Einstein frame, while the quantities with tilde denote those in the physical frame.

By varying the total action $S$ with respect to $g_{*\mu\nu}$, one can obtain the field equation
for the tensor field;
\begin{equation}
 G_{*\mu\nu} = 8\pi G_*T_{*\mu\nu} + 2\left(\varphi_{,\mu}\varphi_{,\nu}
    - \frac{1}{2}g_{*\mu\nu}g_*^{\alpha\beta}\varphi_{,\alpha}\varphi_{,\beta}\right), \label{basic1}
\end{equation}
where $T_{*\mu\nu}$ is the energy-momentum tensor in the Einstein frame, which is associated with that in the physical frame as
\begin{equation}
  T_*^{\mu\nu} \equiv \frac{2}{\sqrt{-g_*}}\frac{\delta S_m}{\delta g_{*\mu\nu}}
     = A^6(\varphi)\tilde{T}^{\mu\nu}.
\end{equation}
Similarly, by varying $S$ with respect to $\varphi$, one obtains the field equation for the scalar field;
\begin{equation}
 \dalm_*\varphi = -4\pi G_*\alpha(\varphi)T_*, \label{basic2}
\end{equation}
where $\dalm_*$ denotes the d'Alembertian of $g_{*\mu\nu}$, and $\alpha(\varphi)$ and $T_*$ are defined as
\begin{gather}
  \alpha(\varphi) \equiv \frac{d\ln A(\varphi)}{d\varphi}, \\
  T_* \equiv T_{*\mu}^{\ \ \mu} = T_*^{\mu\nu}g_{*\mu\nu}.
\end{gather}
We should remark that the scalar-tensor gravity for $\alpha(\varphi)=0$ reduces to the general relativity.
In addition to Eqs. (\ref{basic1}) and (\ref{basic2}), one gets another equation from the energy-momentum conservation law, i.e., $\tilde{\nabla}_\mu\tilde{T}^{\mu}_{\ \nu}=0$, such as
\begin{equation}
  \nabla_{*\mu}T^\mu_{*\nu} = \alpha(\varphi)T_*\nabla_{*\nu}\varphi. \label{basic3}
\end{equation}
Especially, in this article, we adopt the same form of conformal factor $A(\varphi)$ as in Ref. \cite{Damour1993}, i.e.,
\begin{equation}
  A(\varphi) = e^{\beta\varphi^2/2},
\end{equation}
where $\beta$ is a real constant. Thus, $\alpha(\varphi)$ can be written as $\alpha(\varphi)=\beta\varphi$, and the scalar-tensor gravity with $\beta=0$ corresponds to the general relativity.

\subsection{Nonrotating Relativistic Stellar Models in Scalar-Tensor Gravity}
\label{sec:II-2}

The equilibrium configurations of non-rotating relativistic stars in the scalar-tensor gravity have been calculated in Refs. \cite{Damour1992,Harada1998,Sotani2004}. The metric describing the static, spherically symmetric relativistic stars can be written as
\begin{align}
  ds_*^2 &= g_{*\mu\nu}dx^\mu dx^\nu \nonumber \\
      &= -e^{2\Phi(r)}dt^2 + e^{2\Lambda(r)}dr^2+r^2(d\theta^2 + \sin^2\theta\,d\phi^2),
\end{align}
where $e^{-2\Lambda}=1-2\mu(r)/r$, and $\mu(r)$ is corresponding to the mass function. Therefore, the physical metric is
\begin{align}
  d\tilde{s}^2 &= \tilde{g}_{\mu\nu}dx^\mu dx^\nu \nonumber \\
      &= -A^2e^{2\Phi}dt^2 + A^2e^{2\Lambda}dr^2+A^2r^2(d\theta^2 + \sin^2\theta\,d\phi^2).
\end{align}
We assume that the stellar matter is a perfect fluid, i.e., $\tilde{T}_{\mu\nu}=\left(\tilde{\rho}+\tilde{P}\right)\tilde{U}_\mu\tilde{U}_\nu+\tilde{P}\tilde{g}_{\mu\nu}$, where $\tilde{\rho}$, $\tilde{P}$, and $\tilde{U}_\mu$ are the total energy density, the pressure, and the four-velocity of the fluid in physical frame, respectively.

Using Eqs. (\ref{basic1}), (\ref{basic2}), and (\ref{basic3}), one can obtain the Tolman-Oppenheimer-Volkoff (TOV) equations in the scalar-tensor gravity \cite{Damour1992,Harada1998}, such as
\begin{align}
  \mu' &= 4\pi G_* r^2 A^4\tilde{\rho} + \frac{1}{2}r(r-2\mu)\Psi^2, \\
  \Phi' &= 4\pi G_* \frac{r^2A^4\tilde{P}}{r-2\mu} + \frac{1}{2}r\Psi^2 + \frac{\mu}{r(r-2\mu)}, \\
  \varphi' &= \Psi, \\
  \Psi' &= 4\pi G_* \frac{rA^4}{r-2\mu}\left[\alpha\left(\tilde{\rho}-3\tilde{P}\right) + r\left(\tilde{\rho}-\tilde{P}\right)\Psi\right]
       - \frac{2(r-\mu)}{r(r-2\mu)}\Psi, \\
  \tilde{P}' &= -\left(\tilde{\rho}+\tilde{P}\right)\left[\Phi'+\alpha\Psi\right],
\end{align}
where the prime denotes a derivative with respect to $r$. In order to close the system of equations, one needs to prepare the relationship between $\tilde{\rho}$ and $\tilde{P}$, i.e., equation of state (EOS). In this article, we adopt the polytropic EOS as in Refs. \cite{Damour1992,Harada1998,Sotani2004};
\begin{align}
  \tilde{P} &= Kn_0m_b\left(\frac{\tilde{n}}{n_0}\right)^\Gamma, \\
  \tilde{\rho} &= \tilde{n}m_b+ \frac{\tilde{P}}{\Gamma-1}, 
\end{align}
where $\tilde{n}$ is the baryon number density in the physical frame, while $m_b$ and $n_0$ are some constants given by $m_b=1.66\times 10^{-24}$ g and $n_0=0.1$ fm$^{-3}$. In order to fit this EOS to the tabulated data of realistic EOS known as EOS A \cite{eosA} and EOS II \cite{eosII}, we especially adopt $\Gamma=2.46$ and $K=0.00936$ for EOS A and $\Gamma=2.34$ and $K=0.0195$ for EOS II.

The stellar equilibrium can be determined as follows. With the boundary conditions at the stellar center as $\mu(0)=0$, $\Phi(0)=\Phi_c$, $\tilde{\rho}(0)=\tilde{\rho}_c$, $\varphi(0)=\varphi_c$, and $\Psi(0)=0$, the TOV equations with EOS are integrated outward and the stellar surface is determined at $\tilde{P}=0$ as $R=A(\varphi_s)r_s$, where $\varphi_s$ and $r_s$ denote the values of $\varphi$ and $r$ at $\tilde{P}=0$. For $r\ge r_s$, the TOV equations with $\tilde{\rho}=\tilde{P}=0$ are integrated outward. And then, the central values of $\Phi_c$ and $\varphi_c$ can be determined so that the calculated solutions should agree with the asymptotic behaviors as
\begin{align}
  g_{*\mu\nu} &= \eta_{\mu\nu} + \frac{2M_{\rm ADM}}{r} \delta_{\mu\nu} + {\cal O}\left(\frac{1}{r^2}\right), \\
  \varphi(r) &= \varphi_0 + \frac{Q}{r} + {\cal O}\left(\frac{1}{r^2}\right),
\end{align}
where $\eta_{\mu\nu}$, $M_{\rm ADM}$, $\varphi_0$, and $Q$ are corresponding to the Minkowskian metric, the ADM mass, the cosmological value of scalar field, and the total scalar charge, respectively \cite{qs}. In practice, with the matching conditions at the stellar surface between the interior and exterior solutions, one can determine the values of $M_{\rm ADM}$, $\varphi_0$, and $Q$ as
\begin{gather}
 M_{\rm ADM} = \frac{r_s^2 \Phi_s'}{G_*}\left(1-\frac{2\mu_s}{r_s}\right)^{1/2}
      \exp\left[-\frac{\Phi'_s}{\sqrt{(\Phi_s')^2+\Psi_s^2}}\,
      {\rm arctanh}\left(\frac{\sqrt{(\Phi_s')^2+\Psi_s^2}}{\Phi_s'+1/r_s}\right)\right],  \\
 \varphi_0 = \varphi_s + \frac{\Psi_s}{\sqrt{(\Phi_s')^2+\Psi_s^2}} \,
      {\rm arctanh}\left[\frac{\sqrt{(\Phi_s')^2+\Psi_s^2}}{\Phi_s'+1/r_s}\right],  \\
 Q = -\frac{\Psi_s}{\Phi_s'}M_{\rm ADM},
\end{gather}
where the quantities with the subscript $s$ denote their values at $r=r_s$ \cite{Damour1993}. At last, choosing the values of two constants $\beta$ and $\varphi_0$, the stellar models become one parameter family with $\tilde{\rho}_c$ (or $M_{\rm ADM}$). Especially, in this article, we adopt $\varphi_0=0.01$ as a example for $\varphi_0\ne 0$.

\begin{figure}[htbp]
\begin{center}
\begin{tabular}{cc}
\includegraphics[scale=0.5]{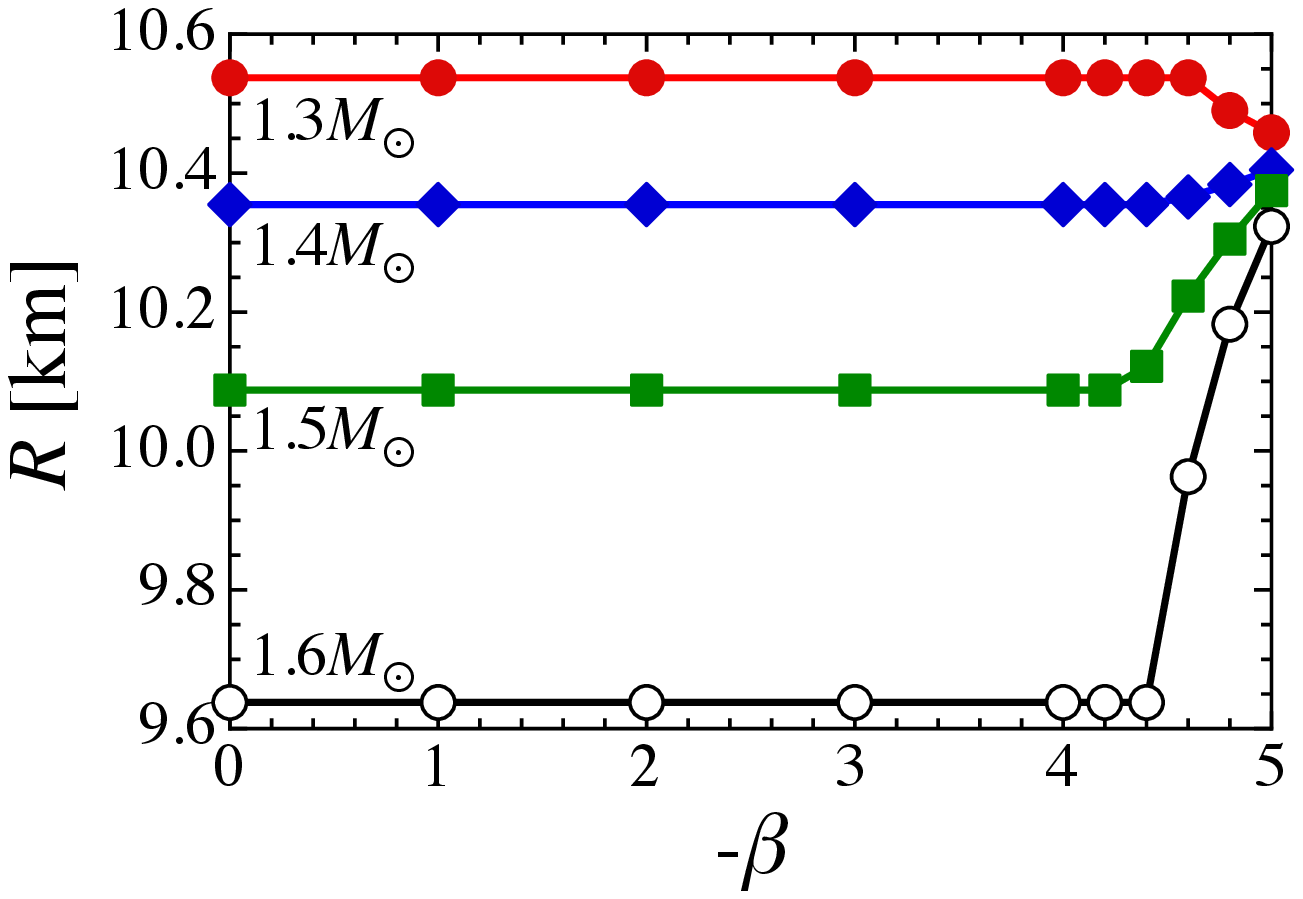} &
\includegraphics[scale=0.5]{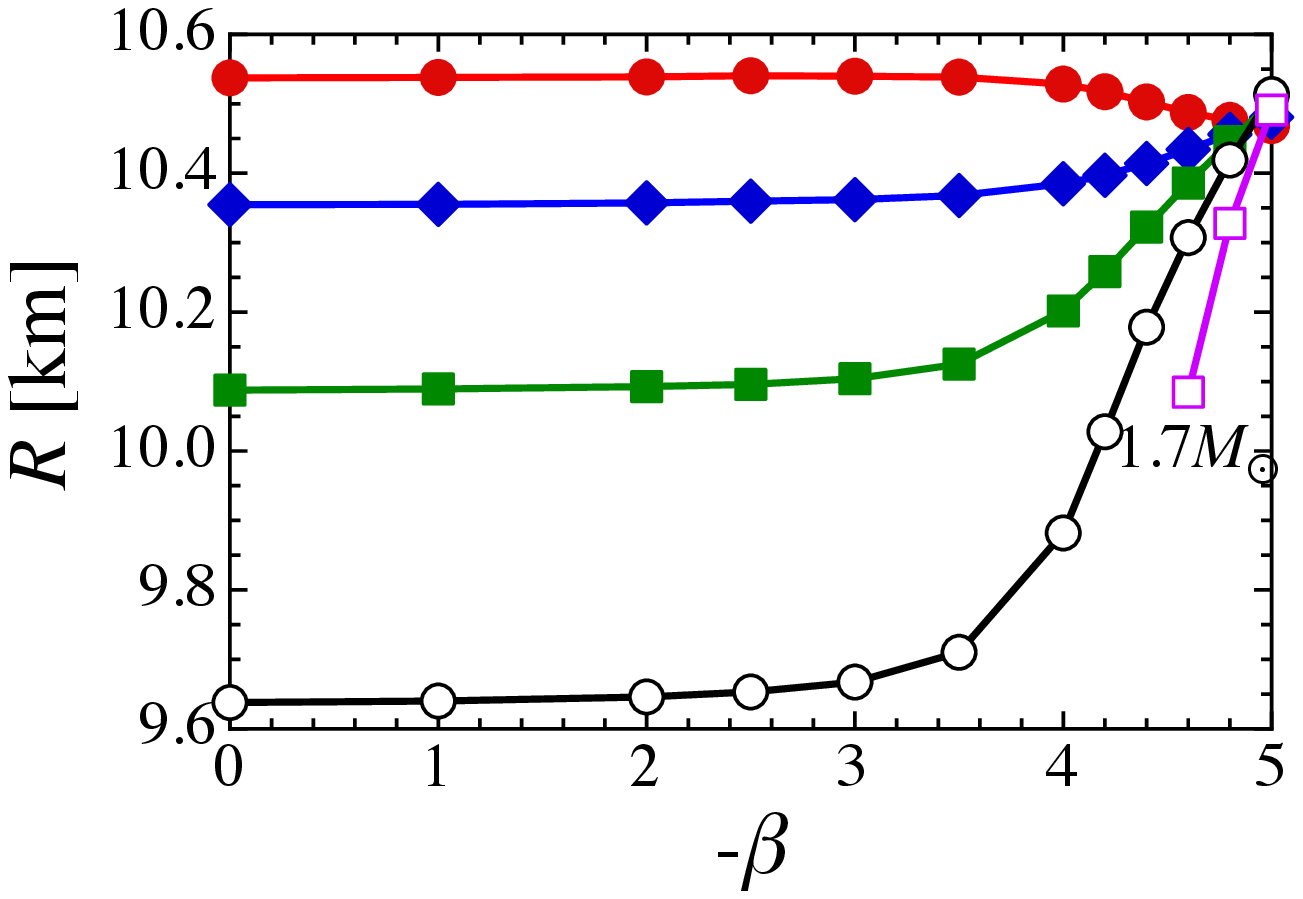} \\
\end{tabular}
\end{center}
\caption{
For EOS A, the stellar radius $R$ is plotted as a function of $-\beta$ with the fixed ADM mass for $\varphi_0=0$ (left panel) and for $\varphi_0=0.01$ (right panel).
}
\label{fig:b-R-A}
\end{figure}

\begin{figure}[htbp]
\begin{center}
\begin{tabular}{cc}
\includegraphics[scale=0.5]{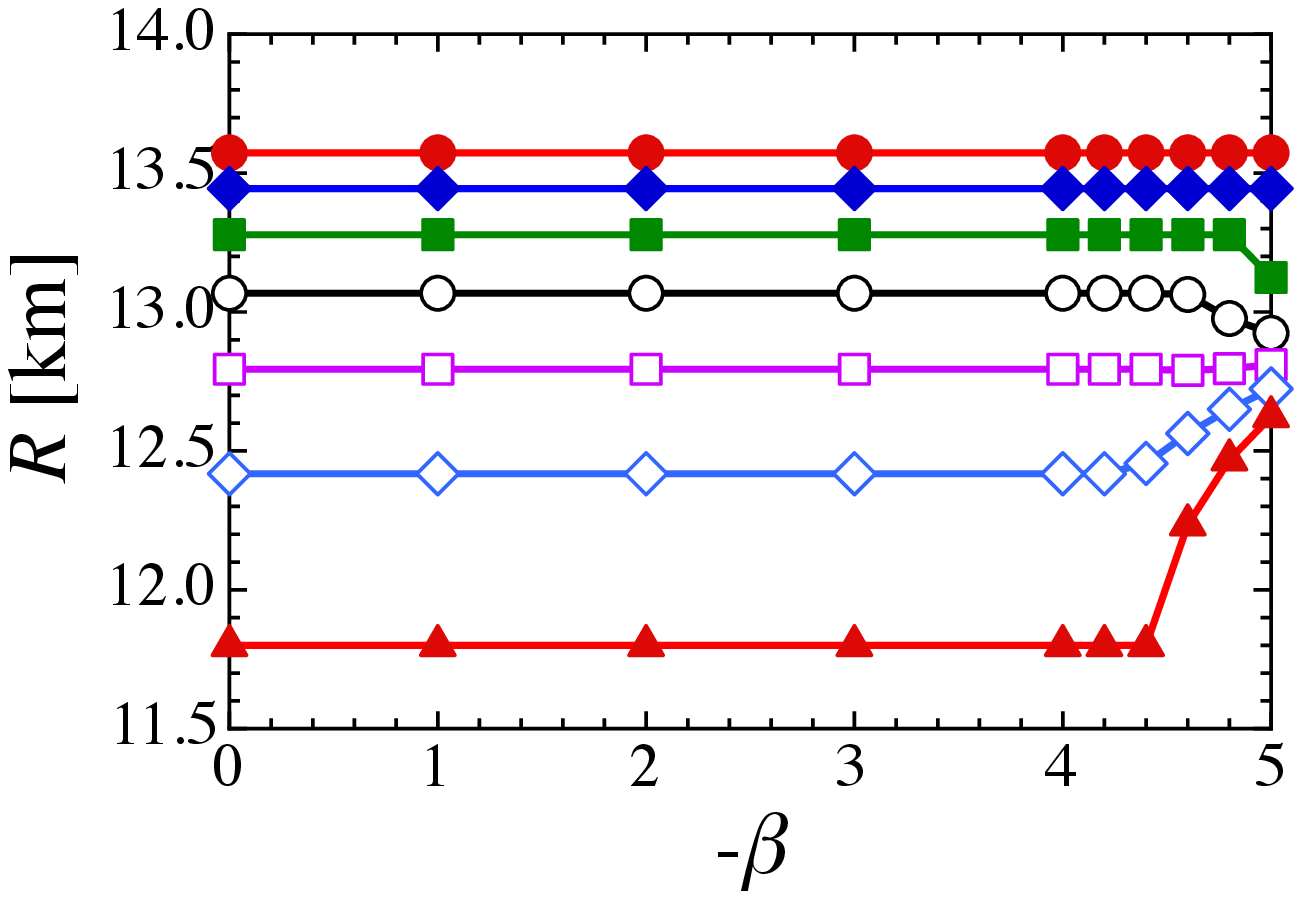} &
\includegraphics[scale=0.5]{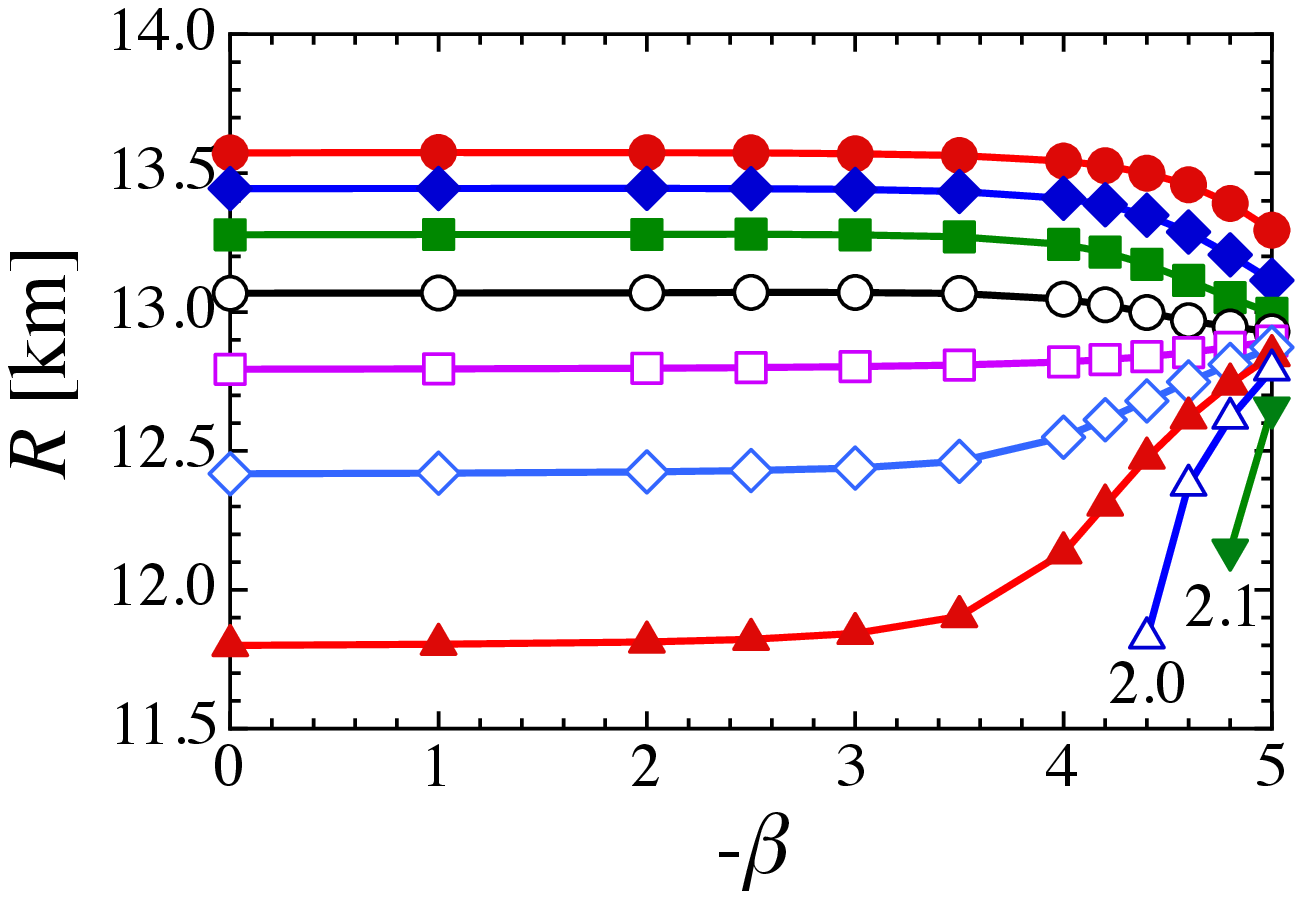} \\
\end{tabular}
\end{center}
\caption{
Similar to Fig. \ref{fig:b-R-A}, but for EOS II. The labels denote the adopted stellar mass in the unit of $M_\odot$, where the seven solid lines from top to bottom correspond to the cases of $M/M_\odot$ = 1.3, 1.4, 1.5, 1.6, 1.7, 1.8, and 1.9.
}
\label{fig:b-R-II}
\end{figure}

We show the dependence of the stellar radius $R$ on the coupling constant $-\beta$ with the fixed ADM mass in Fig. \ref{fig:b-R-A} for EOS A and in Fig. \ref{fig:b-R-II} for EOS II. In both figures, the left and right panels are corresponding to the case of $\varphi_0=0$ and $\varphi_0=0.01$, respectively. From these figures, one can observe the spontaneous scalarization in $R$ at around $\beta\simeq -4.4$ for $\varphi_0=0$, while the stellar radii are changing smoothly for $\varphi_0\ne 0$. Namely, even for $-4.4<\beta <0$, the stellar properties in the scalar-tensor gravity with $\varphi_0\ne 0$ are different from those in the general relativity (for the case of $\beta=0$).

%
%
%

\section{Slowly Rotating Relativistic Stars in Scalar-Tensor Gravity}
\label{sec:III}
\subsection{Rotating Dragging}
\label{sec:III-1}

In order to see the rotational effect around the relativistic stars in the scalar-tensor gravity, we consider a slowly rotating stellar model with a uniform angular velocity, $\tilde{\Omega}$, where the rotational axis is put on $\theta=0$. For this purpose, we assume to keep only the linear effects in the angular velocity, which is the same treatment as in the general relativity \cite{Hartle1967}. With this assumption, the stellar model is still spherically symmetric, because the deformation of stellar shape due to the rotation is of the order $\tilde{\Omega}^2$. Similarly, the rotational effects in the components of metric except for $\tilde{g}_{t\phi}$ become of the order $\tilde{\Omega}^2$. So, the metric describing slowly rotating system in the physical frame can be written as
\begin{equation}
 d\tilde{s}^2 = -A^2e^{2\Phi}dt^2 + A^2e^{2\Lambda}dr^2+A^2r^2(d\theta^2 + \sin^2\theta\,d\phi^2) 
    - 2\omega r^2A^2\sin^2\theta dtd\phi,
\end{equation}
where the last term appears due to the rotational effect. Hereafter, in order to specify the rotational effects, the deviation from the non-rotating stellar model will be expressed by using variables with $\delta$, such as $\delta\tilde{g}_{t\phi}=-\omega r^2A^2\sin^2\theta$. The deviations of the pressure, density, and scalar field due to the rotation are also of the order ${\cal O}(\tilde{\Omega}^2)$, i.e., $\delta\tilde{P}={\cal O}(\tilde{\Omega}^2)$, $\delta\tilde{\rho}={\cal O}(\tilde{\Omega}^2)$, and $\delta\varphi={\cal O}(\tilde{\Omega}^2)$, because those properties should behave in the same way under a reversal in the direction of rotation. With such a ordering, one can show that $\delta\tilde{g}_{\mu\nu}=A^2\delta g_{*\mu\nu}$ and $\delta T_{*\mu\nu}=A^2\delta\tilde{T}_{\mu\nu}$ within the linear order of $\tilde{\Omega}$. Additionally, the deviation of fluid velocity in the physical frame can be described as
\begin{equation}
  \delta \tilde{U}^\mu = \left(0,0,0,\tilde{\Omega}\tilde{U}^t\right),
\end{equation}
where $\tilde{U}^t$ denotes the $t$-component of four velocity for non-rotating stellar model, i.e., $\tilde{U}^t=e^{-\Phi}/A$. Consequently, the nonzero component of the energy-momentum tensor of the order $\tilde{\Omega}$ is only $\delta\tilde{T}_{t\phi}$, which are given as
\begin{equation}
  \delta\tilde{T}_{t\phi} = r^2A^2\left(\tilde{\rho}+\tilde{P}\right)
       \left(\omega-\tilde{\Omega}\right)\sin^2\theta - \tilde{P}\omega r^2A^2\sin^2\theta.
\end{equation}

At last, the rotational effect $\omega$ can be determined by solving the following equation;
\begin{equation}
  \delta G_{*\mu\nu} = 8\pi G_*\delta T_{*\mu\nu} - e^{-2\Lambda}\Psi^2\delta g_{*\mu\nu}, \label{delta_G}
\end{equation}
which comes from the Einstein equation (\ref{basic1}). We should remark that another field equation (\ref{basic2}) can not tell us any information about the rotational effect in the order of $\tilde{\Omega}$, because $\delta \varphi\sim{\cal O}(\tilde{\Omega}^2)$, $\delta\alpha\sim{\cal O}(\tilde{\Omega}^2)$, and $\delta T_*\sim{\cal O}(\tilde{\Omega}^2)$. One can show that only the $(t,\phi)$ component in Eq. (\ref{delta_G}) becomes a nontrivial equation, which can be written down as
\begin{align}
  \omega'' + \left(\frac{4}{r} - \Phi' - \Lambda'\right)\omega' &- 2\left[\Phi'' + \left(\Phi' - \Lambda'\right)\left(\Phi' + \frac{1}{r}\right)
       + \Psi^2\right]\omega + \frac{1}{r^2}e^{2\Lambda}\left(\omega_{,\theta\theta} 
       + 3\omega_{,\theta} \frac{\cos\theta}{\sin\theta}\right) \nonumber \\
       &+ 16\pi G_* A^4 e^{2\Lambda}\left[\tilde{P}\omega + \left(\tilde{\rho} 
       + \tilde{P}\right)\left(\tilde{\Omega}-\omega\right)\right] = 0.  \label{delta_G03}
\end{align}
In general, $\omega\sim{\cal O}(\tilde{\Omega})$ can be expressed as $\omega(r,\theta) = -\omega(r)/\sin\theta\,\partial_\theta P_\ell(\cos\theta)$, where $P_\ell$ is the Legendre polynomial of order $\ell$. With this expression, the bracket of the fourth term in Eq. (\ref{delta_G03}) can be reduced to
\begin{equation}
  \omega_{,\theta\theta} + 3\omega_{,\theta}\frac{\cos\theta}{\sin\theta}
       = (\ell+2)(\ell-1)\omega\frac{\partial_\theta P_\ell}{\sin\theta}.
\end{equation}
Furthermore, in order to reproduce the general relativity limit of the scalar-tensor gravity, i.e., $\beta\to 0$, we especially adopt the case for $\ell=1$ in this article. In practice, one can not adopt an arbitrary value of $\ell$ except for $\ell=1$ to construct the distribution of $\omega$ which satisfies the regularity condition at the stellar center \cite{Hartle1967}. Finally, the equation to solve becomes
\begin{align}
  \omega'' + \left(\frac{4}{r} - \Phi' - \Lambda'\right)\omega' &- 2\left[\Phi'' + \left(\Phi' - \Lambda'\right)\left(\Phi' + \frac{1}{r}\right)
       + \Psi^2\right]\omega + 16\pi G_* A^4 e^{2\Lambda}\left[\tilde{P}\omega + \left(\tilde{\rho} 
       + \tilde{P}\right)\left(\tilde{\Omega} - \omega\right)\right] = 0. \label{omega}
\end{align}
Note that Eq. (\ref{omega}) for $\beta=0$ agrees with the well-known equation describing the frame dragging in the general relativity \cite{Hartle1967}. 

\subsection{Numerical Results}
\label{sec:III-2}

\begin{figure}[htbp]
\begin{center}
\includegraphics[scale=0.5]{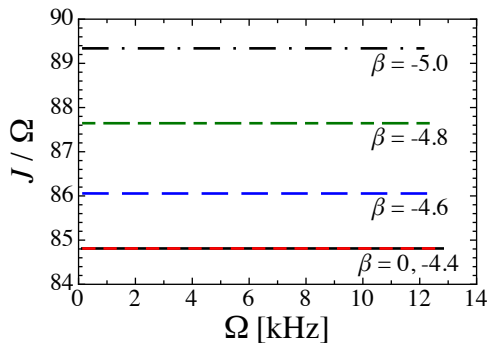} 
\end{center}
\caption{Dependence of $J/\tilde{\Omega}$ on the angular velocity $\tilde{\Omega}$ for the stellar models with EOS A and with $M_{\rm ADM}=1.4M_\odot$ and $\varphi_0=0$, where the solid line corresponds to the case in the general relativity, while the broken lines correspond to the results in the scalar-tensor gravity with different values of $\beta$. 
}
\label{fig:Omega-I-0}
\end{figure}

\begin{figure}[htbp]
\begin{center}
\includegraphics[scale=0.5]{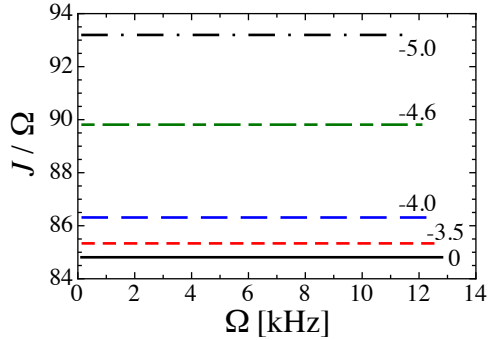} 
\end{center}
\caption{Similar to Fig. \ref{fig:Omega-I-0}, but for $\varphi_0=0.01$. The labels attached on each line denote values of $\beta$.
}
\label{fig:Omega-I-1}
\end{figure}

In order to determine the rotational effect $\omega(r)$ with Eq. (\ref{omega}), one needs to prepare the boundary conditions at the stellar center and the spatial infinity, i.e., the regularity condition at $r=0$ and the asymptotic flatness far from the central object. Since one can expand $\omega$ as $\omega(r)=\omega_{c0}+\omega_{c2}r^2+\cdots$ in the vicinity of stellar center, Eq. (\ref{omega}) is integrated outward from $r=0$ with the boundary conditions of $\omega(0)=\omega_{c0}$ and $\omega'(0)=0$. Then, one can find the exact value of $\omega_{c0}$ in such a way that the solution $\omega(r)$ should satisfy the asymptotic behavior, 
i.e., 
\begin{equation}
  \omega(r) = \frac{2J}{r^3} + {\cal O}\left(\frac{1}{r^4}\right), \label{eq:AB}
\end{equation}
where $J$ is some constant. In practice, we adopt the boundary condition of $1+3\omega/(\omega' r)=0$, which can be derived from Eq. (\ref{eq:AB}), at the numerical boundary far from the central object.
In the general relativity, the distribution of $\omega(r)$ outside the star can be analytically written down as $\omega(r)=2J/r^3$, where the constant $J$ corresponds to the total angular momentum of the central object, and $J$ is related to the angular velocity for slow rotation as 
\begin{equation}
  J = I\tilde{\Omega}. \label{eq:II}
\end{equation}
Here, the constant of proportionality $I$ corresponds to the relativistic generalization of momentum of inertia for slowly rotating system \cite{Hartle1967}. On the other hand, in the scalar-tensor gravity, we plot the value of $J/\tilde{\Omega}$ as a function of the angular velocity $\tilde{\Omega}$ for the stellar models with EOS A and with $M_{\rm ADM}=1.4M_\odot$ in Fig. \ref{fig:Omega-I-0}  for $\varphi_0=0$ and in Fig. \ref{fig:Omega-I-1} for $\varphi_0\ne 0$. In the both figures, the right endpoints of lines are corresponding to the allowed maximum angular velocity for each stellar model. From these figures, one can see that the value of $J/\tilde{\Omega}$ is independent of $\tilde{\Omega}$ even in the scalar-tensor gravity. That is, Eq. (\ref{eq:II}) can hold even in the scalar-tensor gravity. This is the same results as in the TeVeS \cite{Sotani2010}. We remark that in Fig. \ref{fig:Omega-I-0} the lines for $\beta\ge -4.4$ are completely equivalent to the line for $\beta=0$, because the stellar models with $\varphi_0=0$ for $\beta\ge -4.4$ are the same as those in the general relativity (for $\beta=0$).

%
%

\begin{figure}[htbp]
\begin{center}
\includegraphics[scale=0.5]{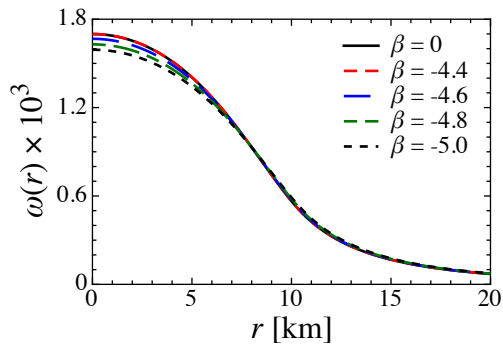} 
\end{center}
\caption{Distribution of $\omega(r)$ for $\varphi_0=0$ with different values of $\beta$, where the adopted stellar model is $M_{\rm ADM}=1.4M_\odot$ and $\tilde{\Omega}=1$ kHz for EOS A.
}
\label{fig:omega-r-0}
\end{figure}

\begin{figure}[htbp]
\begin{center}
\includegraphics[scale=0.5]{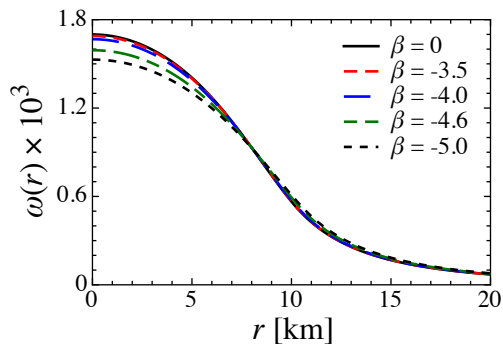} 
\end{center}
\caption{Similar to Fig. \ref{fig:omega-r-0}, but for $\varphi_0=0.01$.
}
\label{fig:omega-r-1}
\end{figure}

The distribution of $\omega(r)$ for the stellar model with EOS A and with $M_{\rm ADM}=1.4M_\odot$ and $\tilde{\Omega}=1$ kHz is plotted in Fig. \ref{fig:omega-r-0} for $\varphi_0=0$ and in Fig. \ref{fig:omega-r-1} for $\varphi_0=0.01$. In the both cases, we can obviously observe the difference of $\omega$ inside the star in the scalar-tensor gravity from the prediction in the general relativity. In fact, compared with the case in the general relativity, one can see the decrease of the central value of $\omega$ in the scalar-tensor gravity up to $6.3\%$ in Fig. \ref{fig:omega-r-0} and $10.0\%$ in Fig. \ref{fig:omega-r-1}. Thus, via the observations such as the stellar oscillations and/or the radiated gravitational waves, one could be possible to probe the gravitational theory in the strong-field regime. Meanwhile, although the deviation outside the star may be almost negligible, there still exists. Now, we should remark the difference between the results in the scalar-tensor gravity and in the TeVeS. Unlike the results in the scalar-tensor gravity, the value of $\omega(r)$ in the TeVeS becomes larger than that in the general relativity in the vicinity of stellar center, and becomes smaller in the outer region (see Fig. 3 in \cite{Sotani2010}). This means that one might be possible to probe the gravitational theory via the observations of binary system composed of the relativistic stars with quite high accuracy.

%
%

\begin{figure}[htbp]
\begin{center}
\begin{tabular}{cc}
\includegraphics[scale=0.5]{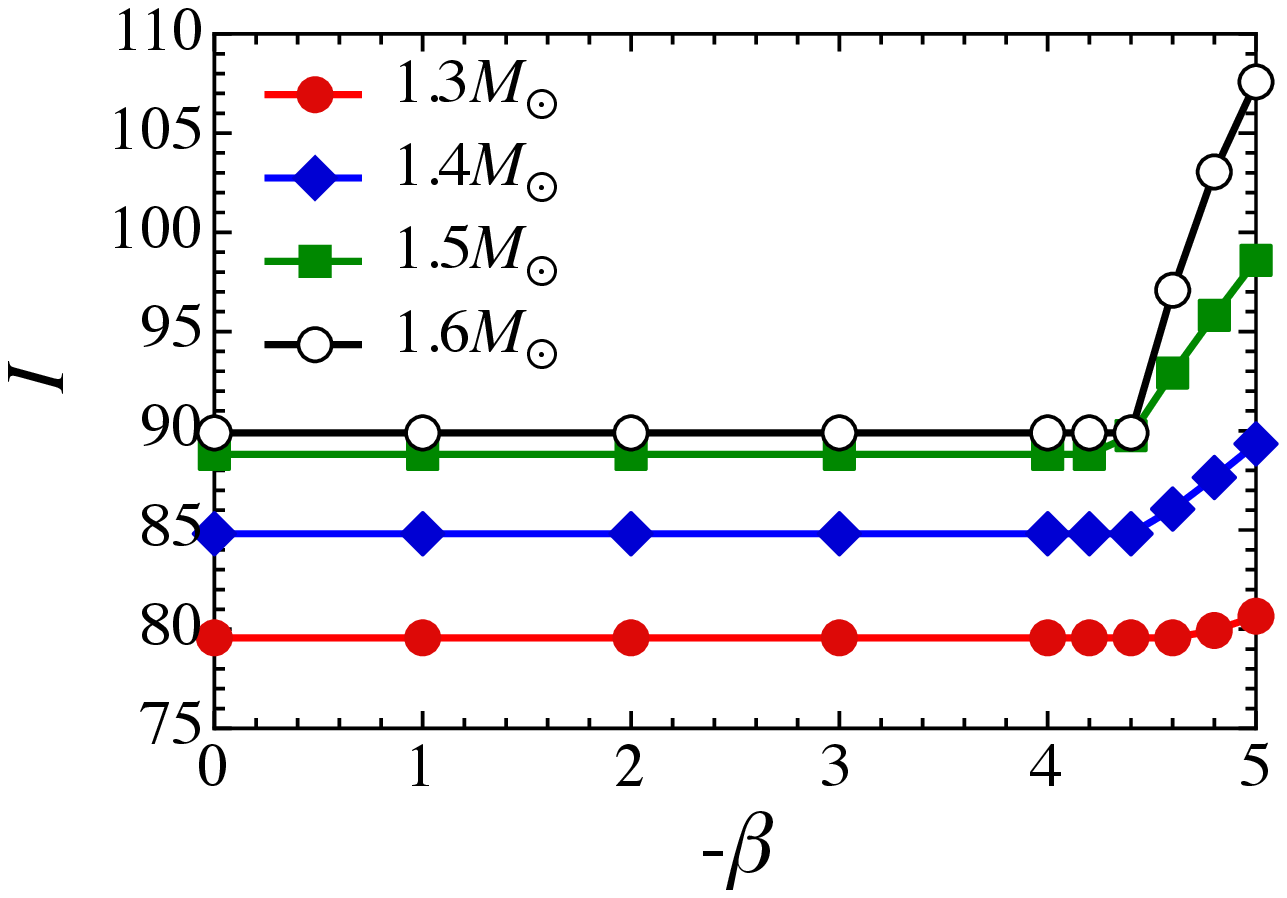} &
\includegraphics[scale=0.5]{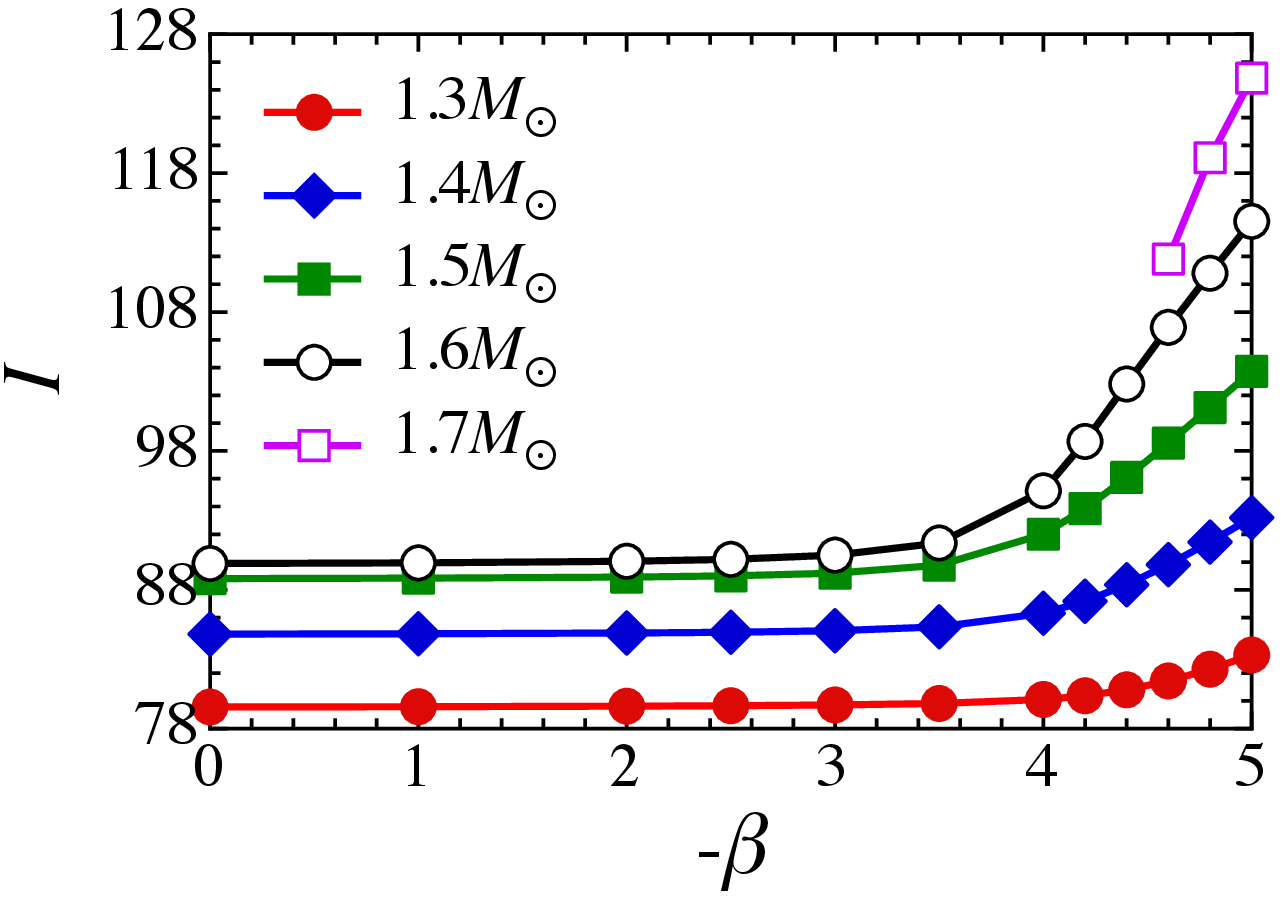} \\
\end{tabular}
\end{center}
\caption{
For EOS A, the momentum of inertia $I$ is plotted as a function of $-\beta$ with the fixed ADM mass for $\varphi_0=0$ (left panel) and for $\varphi_0=0.01$ (right panel).
}
\label{fig:b-I}
\end{figure}

\begin{figure}[htbp]
\begin{center}
\begin{tabular}{cc}
\includegraphics[scale=0.5]{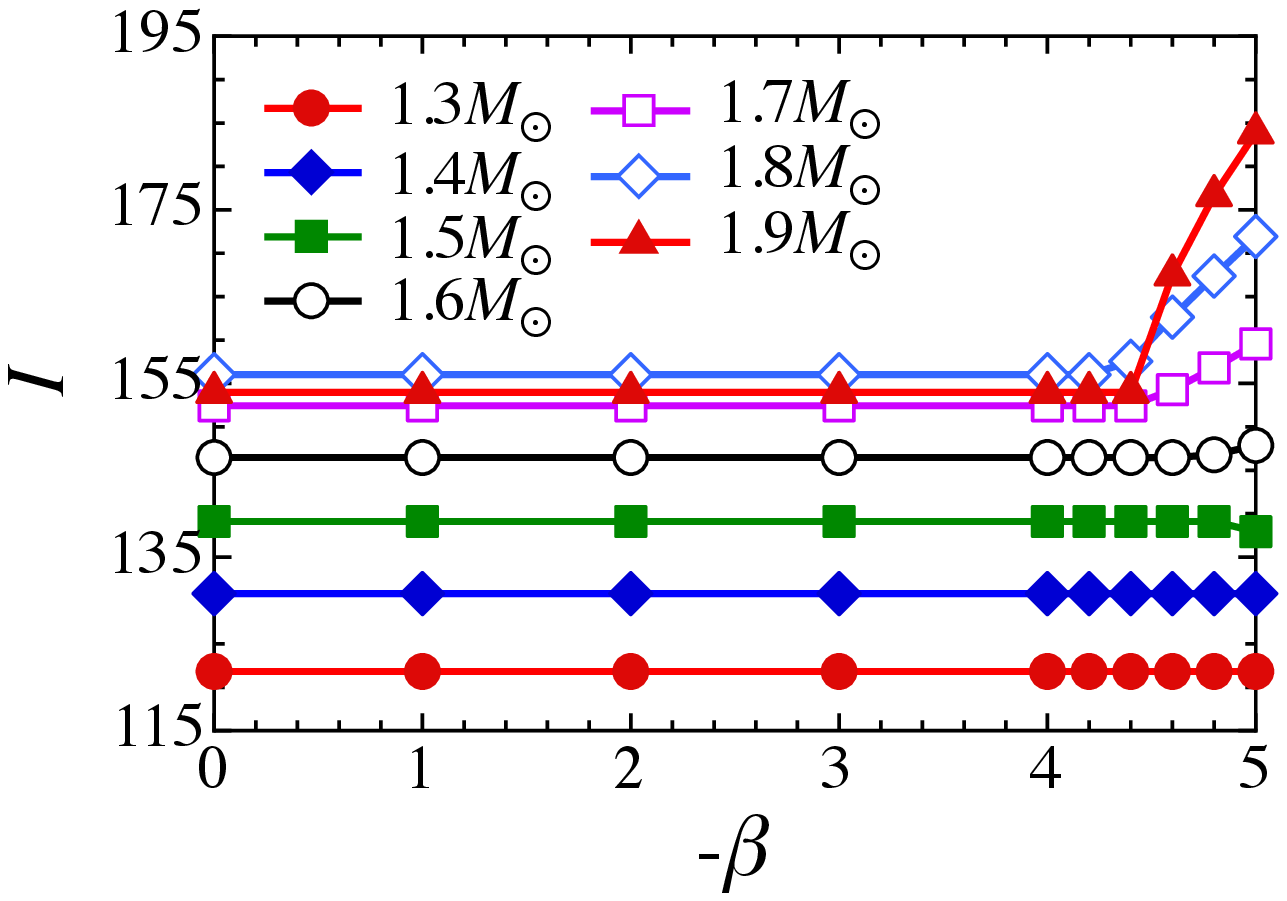} &
\includegraphics[scale=0.5]{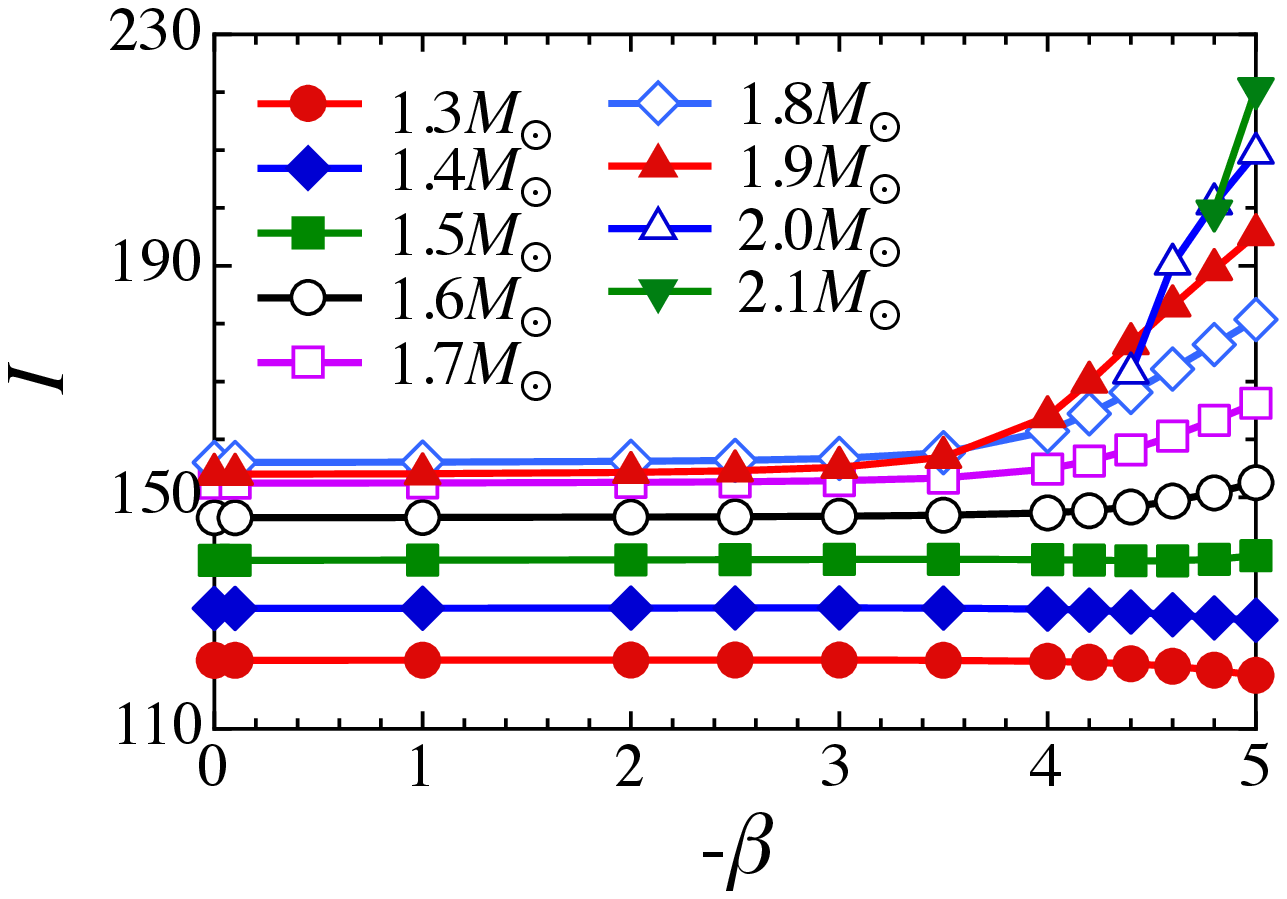} \\
\end{tabular}
\end{center}
\caption{
Similar to Fig. \ref{fig:b-I}, but for EOS II.
}
\label{fig:b-I-II}
\end{figure}

\begin{figure}[htbp]
\begin{center}
\begin{tabular}{cc}
\includegraphics[scale=0.5]{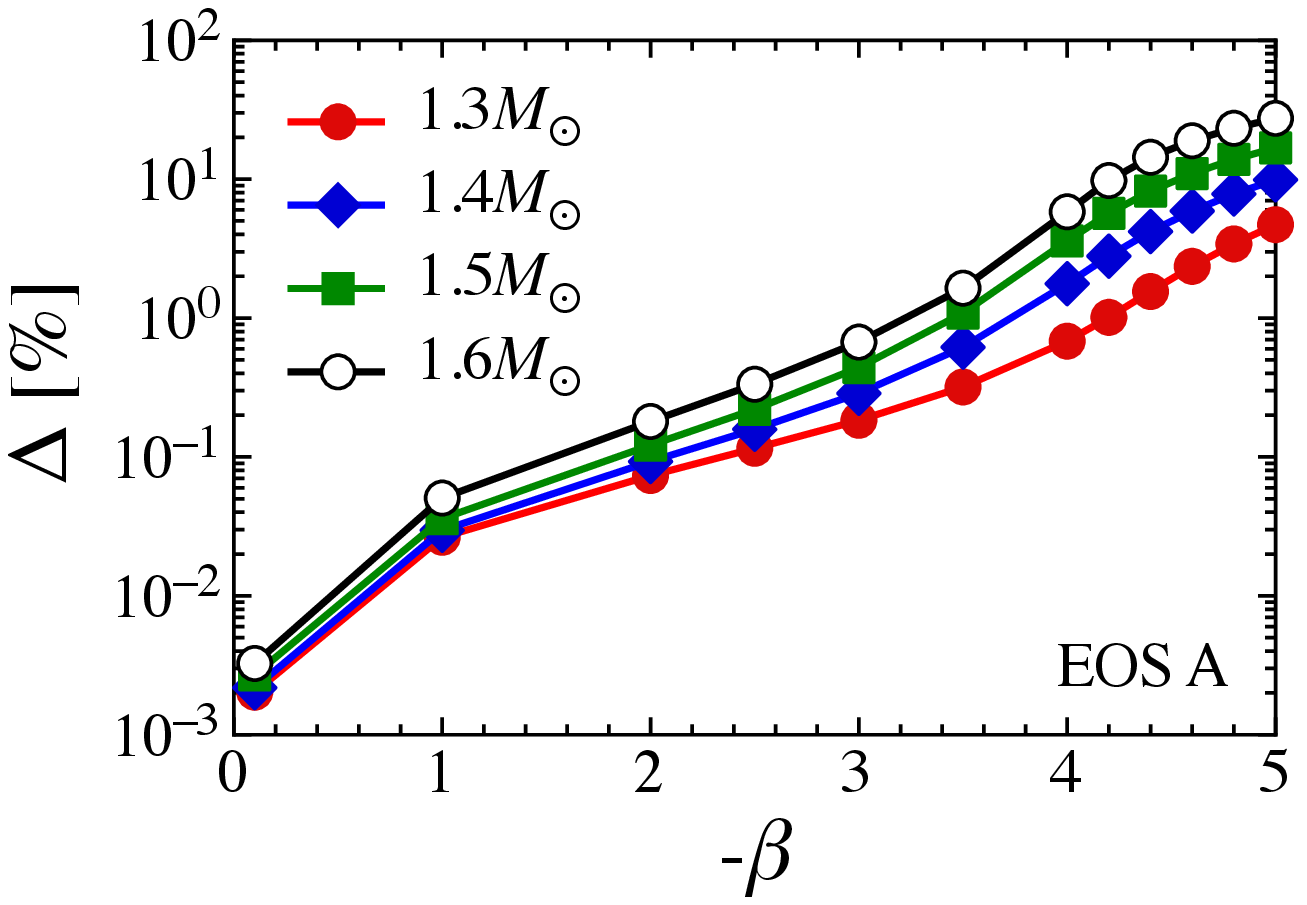} &
\includegraphics[scale=0.5]{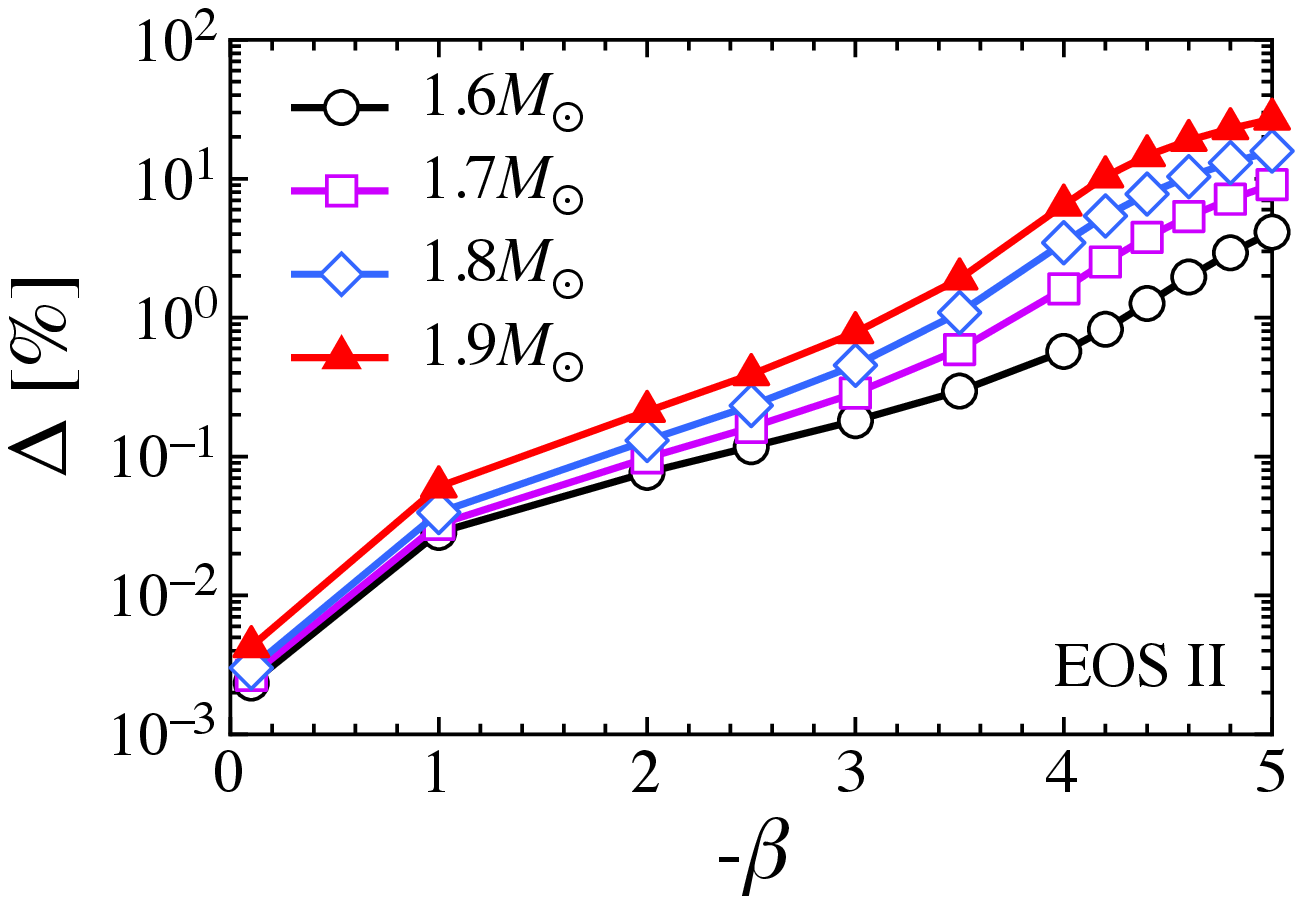} \\
\end{tabular}
\end{center}
\caption{
Relative deviation of $I$ in the scalar-tensor gravity compared with $I$ in the general relativity. The left and right panels correspond to the case of EOS A and EOS II, respectively.
}
\label{fig:b-D}
\end{figure}

Furthermore, we plot the momentum of inertia $I$ as a function of the coupling constant $-\beta$ with the fixed ADM mass in Fig. \ref{fig:b-I} 
for EOS A and in Fig. \ref{fig:b-I-II} for EOS II, where the left and right panels are results for $\varphi_0=0$ and for $\varphi_0=0.01$. From the both figures, it is found that the dependence of $I$ on $-\beta$ becomes stronger, as the stellar mass increases. One can also observe that, compared with the case of $\varphi_0=0$, the case of $\varphi_0\ne 0$ depends strongly on the scalar coupling $\beta$. For example, for EOS A, the value of $I$ in the scalar-tensor gravity becomes up to 1.4\%, 5.3\%, 11.0\%, and 19.7\% larger than that in the general relativity for $M=1.3M_\odot$, $1.4M_\odot$,  $1.5M_\odot$, and  $1.6M_\odot$ in the left panel of Fig. \ref{fig:b-I}, while up to 4.7\%, 9.9\%, 16.8\%, and 27.4\% larger for $M=1.3M_\odot$, $1.4M_\odot$,  $1.5M_\odot$, and  $1.6M_\odot$ in the right panel of Fig. \ref{fig:b-I}. We should emphasize that such a dependence of $I$ on the gravitational theory is completely different from that in the case of TeVeS, i.e., the value of $I$ in the TeVeS becomes smaller than that in the general relativity \cite{Sotani2010}. Additionally, for the case of $\varphi_0=0$ (left panels in Figs. \ref{fig:b-I} and \ref{fig:b-I-II}), only if $\beta$ is less than around $-4.4$, $I$ in the scalar-tensor gravity can deviate from $I$ in the general relativity. Namely, the spontaneous scalarization can be observed in $I$ as well as the other stellar properties. On the other hand, for the case of $\varphi_0\ne 0$ (right panels in Figs. \ref{fig:b-I} and \ref{fig:b-I-II}), one can observe the deviation in $I$ even for $-4.4\lsim \beta <0$. Now, in order to clarify the deviation in $I$ for $\varphi_0=0.01$, we define the relative deviation $\Delta$ as
\begin{equation}
  \Delta = (I_{\rm ST}-I_{\rm GR}) / I_{\rm GR}, 
\end{equation}
where $I_{\rm ST}$ and $I_{\rm GR}$ denote the momentum of inertia defined as Eq. (\ref{eq:II}) in the scalar-tensor gravity and in the general relativity, respectively. The calculted $\Delta$ is plotted as a function of $-\beta$ in Fig. \ref{fig:b-D}, where the left and right panels are corresponding to the results for EOS A and for EOS II. This figure obviously shows the statement mentioned the above, i.e., the deviation from the general relativity becomes larger with more massive stellar model. Meanwhile, comparing the stellar models with same mass but different EOS, one can see that the stellar model with softer EOS becomes larger deviation from the prediction in the general relativity. At last, one can clearly see the deviation from the general relativity even for $-4.4\lsim\beta <0$, which is around $0.01-10$ \%. This deviation might be small, but we still have a chance to probe the gravitational theory in the strong-field regime via the accurate observations
around the relativistic objects. 

%
%
%
%
%

\section{Conclusion}
\label{sec:IV}

In order to examine the rotational effect around the relativistic stars in the scalar-tensor gravity, we consider the slowly rotating relativistic objects with a uniform angular velocity. From the Einstein equations, we derive a second order differential equation governing the frame dragging, and solve it numerically with appropriate boundary conditions. As a result, we find that, similarly in the general relativity, the total angular momentum is proportional to the angular velocity for the slowly rotation. Additionally, we find that the spontaneous scalarization can arise in the rotational effects as well as the other stellar properties for $\varphi_0=0$, where $\varphi_0$ denotes the cosmological value of scalar field. On the other hand, for $\varphi_0\ne 0$, we can observe the deviation from the general relativity in a wide range of the coupling constant $\beta$. Comparing the results in this article with those in the case of TeVeS, we find that the dependence on the gravitational theory is completely different. Especially, we find the obvious difference in the momentum of inertia. That is, compared with the prediction in the general relativity, the momentum of inertia in the TeVeS becomes smaller, while that in the scalar-tensor gravity becomes larger.
Thus, via the astronomical observations around the relativistic stars with high accuracy, one could possible to probe not only the gravitational theory in the strong-field regime, but also the existence of scalar field. At last, we adopt simple stellar models in this article, but to compare with the actual observational data, we need to consider the more realistic stellar models, where we might take into account the magnetic effects \cite{Sotani2007} and the effects of crust and/or the more exotic phase  \cite{crust,HQ}. Considering such additional properties, one could be possible to obtain the further constraint in the theory.

\acknowledgments

This work was supported in part by Grants-in-Aid for Scientific Research on Innovative Areas through No.\ 23105711, No.\ 24105001, and No.\ 24105008 provided by MEXT, by Grant-in-Aid for Young Scientists (B) through No.\ 24740177 provided by JSPS, by the Yukawa International Program for Quark-hadron Sciences, and by the Grant-in-Aid for the global COE program ``The Next Generation of Physics, Spun from Universality and Emergence" from MEXT.





\begin{thebibliography}{999}
\bibitem{rotationinBD}
 Previously, Reizlin also considered the rotational effect in the Brans-Dicke theory \cite{Reizlin1981}, but as mentioned in text, we can not expect the significant deviation in this case from the prediction in the general relativity.
\end{thebibliography}
\end{document}